\documentclass[12pt]{iopart}

\usepackage{iopams}   

\begin{document} 

\jl{1} 
 
\title[Coupling coefficients of SO(n)]{More on coupling coefficients for the
  most degenerate representations of \boldmath{$ SO(n) $}} 

\author{Markus Horme{\ss}\footnote[1]{E-mail:
  hormess@theorie1.physik.uni-erlangen.de}   
  and 
  Georg Junker\footnote[2]{E-mail: 
    junker@theorie1.physik.uni-erlangen.de} }  
 
\address{Institut f\"ur Theoretische Physik, 
Universit\"at Erlangen-N\"urnberg, Staudtstr.\ 7, D-91058 Erlangen, Germany} 
 
\begin{abstract} 
We present explicit closed-form expressions for the general group-theoretical
factor appearing in the $\alpha $-topology of a high-temperature 
expansion of $SO(n)$-symmetric lattice models. This object, which is closely
related to $6j$-symbols for the most degenerate representation of $SO(n)$, is
discussed in detail. 
\end{abstract}
~

\section{Introduction}
In this paper we extend our previous studies \cite{1} on coupling coefficients
for the so-called most degenerate (also named symmetric or class-one)
representations of $SO(n)$.  
These coupling coefficients are important in many fields of
theoretical physics such as atomic and nuclear physics. For example, in
connection with the Jahn-Teller effect an extensive study of particular
$6j$-symbols is due to Judd and coworkers \cite{2}. A detailed study of
isoscalar factors of $SO(n)\supset SO(n-1)$ and related $6j$-coefficients is
due to Ali\v{s}auskas \cite{Sigitas} showing that the $6j$-coefficients of
$SO(n)$ can be expressed in terms of (generalized) $6j$-coefficients of
$SU(2)$. 

Coupling coefficients of the most degenerate representations of $SO(n)$, also
appear as group-theoretical factors in the high-temperature expansion of 
$SO(n)$-symmetric classical lattice models \cite{3,4} such as the
$ XY$-model ($n=2$) and the Heisenberg model ($n=3$). 
In this paper we present new explicit results for the so-called 
$ \alpha$-graph, which contributes with the following group-theoretical factor
to the high-temperature expansion of the free energy of such models
\cite{3,4,1}. 
\begin{equation}
  \fl\label{1}
  \begin{array}{l}
  I_n\equiv I_n(\ell_1,\ell_2,\ell_3 |\ell_4,\ell_5,\ell_6) \\[2mm]
   \mbox{~~}\displaystyle
  :=\!\!\!\int\limits_{SO(n)}\!\!\!\rmd g_1 \!\!\!
    \int\limits_{SO(n)}\!\!\!\rmd g_2 \!\!\!
    \int\limits_{SO(n)}\!\!\!\rmd g_3\,
    {\cal D}^{\ell_1}_{00}(g_1){\cal D}^{\ell_2}_{00}(g_2)
    {\cal D}^{\ell_3}_{00}(g_3){\cal D}^{\ell_4}_{00}(g_2^{-1}g_3)
    {\cal D}^{\ell_5}_{00}(g_3^{-1}g_1){\cal D}^{\ell_6}_{00}(g_1^{-1}g_2)
  \end{array}
\end{equation}
Here $ {\cal D}^{\ell}_{00}(g)$ denotes a particular matrix element (the zonal
spherical function) of the $ \ell$th unitary irreducible class-one
representation of $ SO(n)$, $\ell\in{\mathbb N}_0$, and $\rmd g$ is the
normalised invariant Haar measure on $ SO(n)$. 
For details we refer to our earlier work
\cite{1}. Here we only note the relation of the above integral with the 
$6j$-symbols of $SO(n)$: 
\begin{equation}
  \fl\label{2}
  \begin{array}{rl}
  I_n=&(-1)^{\ell_4+\ell_5+\ell_6}
\left(\begin{array}{ccc}
      \ell_1 &\ell_2 &\ell_3\\ 0 & 0 & 0
      \end{array}\right)_{(n)}
\left(\begin{array}{ccc}
      \ell_1 &\ell_5 &\ell_6\\ 0 & 0 & 0
      \end{array}\right)_{(n)}
\left(\begin{array}{ccc}
      \ell_4 &\ell_2 &\ell_6\\ 0 & 0 & 0
      \end{array}\right)_{(n)}\\
&\times
\left(\begin{array}{ccc}
      \ell_3 &\ell_4 &\ell_5\\ 0 & 0 & 0
      \end{array}\right)_{(n)}
\left\{\begin{array}{ccc}
      \ell_1 &\ell_2 &\ell_3\\ \ell_4 &\ell_5 &\ell_6
      \end{array}\right\}_{(n)}\;,
  \end{array}
\end{equation}
where
\begin{equation}
 \fl \label{3}
  \begin{array}{l}
\left(\begin{array}{ccc}
      \ell_1 &\ell_2 &\ell_3\\ 0 & 0 & 0
      \end{array}\right)_{(n)}^2
:=\displaystyle
\int\limits_{SO(n)}\rmd g\,
{\cal D}^{\ell_1}_{00}(g){\cal D}^{\ell_2}_{00}(g){\cal D}^{\ell_3}_{00}(g)\\
\hspace{25mm}=\displaystyle
\frac{(J+n-3)!}{(n-3)!\,\Gamma^2(n/2)\,\Gamma(J+n/2)}
\prod_{i=1}^3 
\left[\frac{(n-2)!\,\ell_i!\,\Gamma(J-\ell_i+\frac{n-2}{2})}
                         {2\,(\ell_i+n-3)!\,(J-\ell_i)!}\right]
  \end{array}
\end{equation}
denotes the square of a $ 3j$-symbol which vanishes unless 
$J:=(\ell_1+\ell_2+\ell_3)/2$ is a non-negative integer, $J\in{\mathbb N}_0$,
and the $ \ell$'s obey the triangular relation well-known from the case
$ n=3$. This result, in essence, goes back to an earlier one of Vilenkin, ref
\cite{6} equation (6) on page 490. See also the 
work of Ali\v{s}auskas \cite{Sigitas83} and references therein.
A derivation of \eref{3} can be found in \cite{1}, equations (21)-(24), where
also a phase convention for the $3j$-symbol is given.  
This together with an explicit expression for $I_n$ then also leads to
a closed-form expression for the $6j$-symbol which is denoted with curly
brackets in (\ref{2}).  The resulting expressions are indeed similar to those
obtained by Ali\v{s}auskas \cite{Sigitas}.

The purpose of this paper is to derive a rather elementary expression for
the above group integral $I_n(\ell_1,\ell_2,\ell_3 |\ell_4,\ell_5,\ell_6)$
which allows to present, for given but 
arbitrary values of the $\ell$'s and any $n$, explicit results for
(\ref{1}). So far only particular results have been given in the literature. 
For example, for arbitrary $n$ and 
$(\ell_1,\ell_2,\ell_3,\ell_4,\ell_5,\ell_6)=(1,1,2,1,1,2)$ an explicit
expression has been given by Domb \cite{4}, the elementary case $ \ell_4=0 $
can be found in \cite{1}\footnote{
Note that equation (47) in \cite{1} should read  
$\left\{\begin{array}{ccc}
 \ell_1 &\ell_2 &\ell_3\\[-1mm] 0 &\ell_5 &\ell_6 \end{array}\right\}_{(n)}=
 \frac{(-1)^{\ell_1+\ell_2+\ell_3}}
      {\sqrt{d_{\ell_2}d_{\ell_3}}}\,
      \delta_{\ell_2\ell_6}\delta_{\ell_3\ell_5}$
if $\ell_1,\ell_2,\ell_3$ obey the triangular condition and
vanishes otherwise.} 
and, rather recently, some
results have been given for the cases 
where one of the $ \ell$'s equals one or two \cite{5}. 

The remaining part of this paper deals with the derivation of an elementary
expression for $ I_n$, which is given below in \eref{11} in combination with
\eref{7}, \eref{15} and \eref{16}. Together with the above expression for the
$3j$-symbol we have thus also obtained a new elementary expression for the
corresponding $6j$-symbol. We finally present some explicit results for
arbitrary $n$ and $\ell_i\in\{1,2,3,4\}$ after briefly discussing the symmetry
properties of $ I_n$. Our result will also be compared with that of
Ali\v{s}auskas \cite{Sigitas}. 

\section{Explicit integration of (\ref{1})}
In this section  we will make extensive use of our previous
results \cite{1}. In the following if we refer to equations of ref \cite{1} we
will add the superscript 1 to the equation number. For example, $(18)^1$
refers to eq (18) of \cite{1} which shows that the zonal spherical functions
can be expressed in terms of Gegenbauer polynomials. In fact, using this
relation the integral \eref{1} may be rewritten as follows:
\begin{equation}
 \fl \label{5}
 \begin{array}{rl}
I_n=&\displaystyle
\left[\prod_{i=1}^6 
          \frac{\ell_i!\,(n-3)!}{(\ell_i+n-3)!}\right]
    \int\limits_{S^{n-1}} \frac{\rmd^{n-1}\bi{e}_1}{|S^{n-1}|}\,
    \int\limits_{S^{n-1}} \frac{\rmd^{n-1}\bi{e}_2}{|S^{n-1}|}\,
    \int\limits_{S^{n-1}} \frac{\rmd^{n-1}\bi{e}_3}{|S^{n-1}|}\\[6mm]
  &\displaystyle\times
    C_{\ell_1}^{(n-2)/2 }(\bi{a}\cdot\bi{e}_1)\,
    C_{\ell_2}^{(n-2)/2 }(\bi{a}\cdot\bi{e}_2)\,
    C_{\ell_3}^{(n-2)/2 }(\bi{a}\cdot\bi{e}_3)\\[2mm]
  &\displaystyle\times
    C_{\ell_4}^{(n-2)/2 }(\bi{e}_2\cdot\bi{e}_3)\,
    C_{\ell_5}^{(n-2)/2 }(\bi{e}_3\cdot\bi{e}_1)\,
    C_{\ell_6}^{(n-2)/2 }(\bi{e}_1\cdot\bi{e}_2)\;.
 \end{array}
\end{equation}
Here and in the following we will use the same notation as in
\cite{1}. Denoting with $\theta_i$ the polar angle of the unit vector
$\bi{e}_i\in S^{n-1}$ we have  
$\bi{e}_i=(\sin\theta_i\bi{f}_i,\cos\theta_i)$ with $\bi{f}_i\in S^{n-2}$.
Using $\bi{a}\cdot\bi{e}_i=\cos\theta_i$ and the addition theorem
for Gegenbauer polynomials \cite{7}
\begin{equation}
 \fl \label{6}
 \begin{array}{rl}
C_{\ell}^{n/2-1}(\bi{e}_i\cdot\bi{e}_j) =&
 \displaystyle
 \sum_{m=0}^\ell a(n,\ell,m)\,
 \sin^m\theta_i\, C_{\ell-m}^{m+n/2-1}(\cos\theta_i)\,
 \sin^m\theta_j \,C_{\ell-m}^{m+n/2-1}(\cos\theta_j)\\[4mm]
 &\displaystyle\times C_{m}^{(n-3)/2 }(\bi{f}_i\cdot\bi{f}_j)\;,
 \end{array}
\end{equation}
where we have set
\begin{equation}
   \fl \label{7}
a(n,\ell,m):=\frac{2^{2m}(n-4)!\,(\ell-m)!\,\Gamma^2(m+n/2-1)}
                 {(\ell+m+n-3)!\,\Gamma^2(n/2-1)}\,(2m+n-3)\;,
\end{equation}
the above integrations can be factorised into those over the polar angles and
the remaining integrals over $S^{n-2}$. For this we have also to make use of
$(39)^1$ in the form 
\begin{equation}
\fl  \label{8}
  \int\limits_{S^{n-1}}\frac{\rmd^{n-1}\bi{e}}{|S^{n-1}|}\;(\cdot)=
  \frac{\Gamma(n/2)}{\sqrt{\pi}\,\Gamma((n-1)/2)}
  \int\limits_0^\pi\rmd\theta\sin^{n-2}\theta
  \int\limits_{S^{n-2}}\frac{\rmd^{n-2}\bi{f}}{|S^{n-2}|}\;(\cdot)\,.
\end{equation}

The part of \eref{5} which involves the $\bi{f}$-integrations reads
($m_i=0,1,\ldots,\ell_{3+i}$) 
\begin{equation}
 \fl \label{9}
 \begin{array}{rl}
F_n:=&\displaystyle 
  \int\limits_{S^{n-2}}\frac{\rmd^{n-2}\bi{f}_1}{|S^{n-2}|}
  \int\limits_{S^{n-2}}\frac{\rmd^{n-2}\bi{f}_2}{|S^{n-2}|}
  \int\limits_{S^{n-2}}\frac{\rmd^{n-2}\bi{f}_3}{|S^{n-2}|}\\[3mm]
  &\times C_{m_1}^{(n-3)/2 }(\bi{f}_2\cdot\bi{f}_3)\
          C_{m_2}^{(n-3)/2 }(\bi{f}_3\cdot\bi{f}_1)\
          C_{m_3}^{(n-3)/2 }(\bi{f}_1\cdot\bi{f}_2)\\[4mm]
 =&\displaystyle
   \prod_{i=1}^3\left[\frac{(m_i+n-4)!}{m_i!\,(n-4)!}\right]\\
  &\displaystyle\times
    \int\limits_{SO(n-1)}\rmd h_1
    \int\limits_{SO(n-1)}\rmd h_2
    \int\limits_{SO(n-1)}\rmd h_3 \;
    D^{m_1}_{00}(h_2^{-1}h_3)\
    D^{m_2}_{00}(h_3^{-1}h_1)\
    D^{m_3}_{00}(h_1^{-1}h_2)
 \end{array}
\end{equation}
where $D^m_{00} $ denotes zonal spherical functions of the subgroup
$ SO(n-1)$. These group integrations are easily performed via the orthogonality
relation for the $ SO(n-1)$ matrix elements $D^m_{00}$, cf $(12)^1$. As a
consequence all $m_i $'s have to be equal, 
$ m\equiv m_1=m_2=m_3 =0,1,2,\ldots,\min\{\ell_4,\ell_5,\ell_6\}$, 
and the result reads
\begin{equation}
 \fl \label{10}
  F_n=\sum_{m=0}^{\min\{\ell_4,\ell_5,\ell_6\}}
    \delta_{mm_1}\delta_{mm_2}\delta_{mm_3}\,
    \frac{(m+n-4)!}{m!\,(n-4)!}
    \left(\frac{n-3}{2m+n-3}\right)^2\;.
\end{equation}

With the help of this result we are now able to put the quantity of our
interest into the form
\begin{equation}
  \fl\label{11}
   \begin{array}{rl}
I_n=&\displaystyle
\left[\prod_{i=1}^6 
          \frac{\ell_i!\,(n-3)!}{(\ell_i+n-3)!}\right]
\left(\frac{\Gamma(n/2)}{\sqrt{\pi}\,\Gamma((n-1)/2)}\right)^3\\[4mm]
  &\displaystyle\times
   \sum_{m=0}^{\min\{\ell_4,\ell_5,\ell_6\}}
    \frac{(m+n-4)!}{m!\,(n-4)!}
    \left(\frac{n-3}{2m+n-3}\right)^2
\left[\prod_{i=4}^6 a(n,\ell_i,m)\right]\\[3mm]
  &\displaystyle\hspace{20mm}\times
   G_n(\ell_1,\ell_5,\ell_6,m)\,G_n(\ell_2,\ell_4,\ell_6,m)\,
   G_n(\ell_3,\ell_4,\ell_5,m)
        \end{array}
\end{equation}
and thus have reduced it to three elementary integrals of the type
\begin{equation}
 \fl \label{12}
 \begin{array}{rl}
  G_n(j_1,j_2,j_3,m):=&\displaystyle
   \int\limits_0^\pi\rmd\theta \,\sin^{2m+n-2}\theta\\
  &\displaystyle\times
  C_{j_1}^{n/2-1}(\cos\theta)\, C_{j_2-m}^{m+n/2-1}(\cos\theta)\,
  C_{j_3-m}^{m+n/2-1}(\cos\theta)\;.
 \end{array}
\end{equation}
This integral is a special case of a class of integrals already studied in
\cite{1}, cf $ (41)^1$, where we have been able to represent such integrals by
three finite sums.  However, because of its
special form we have decided to evaluate \eref{12} in a different way. In
doing so we first recall the recurrence relation \cite{8} for the Gegenbauer
polynomials 
\begin{equation}
  \label{13}
  C_j^\lambda(x)=\frac{\lambda}{j+\lambda}
  \left[ C_j^{\lambda+1}(x)- C_{j-2}^{\lambda+1}(x)\right]\;,
\end{equation}
which is also valid for $ j=0,1$ if we use the convention that Gegenbauer
polynomials with a ``negative degree'' (the lower index) vanish identically.
Iterating this recurrence relation $ m$ times we find
\begin{equation}
  \label{14}
 C_j^{n/2-1}(\cos\theta)=\sum_{k=0}^{\min\{m,[j/2]\}}(-1)^k\,
 b(n,j,k,m)\,C_{j-2k}^{m+n/2-1}(\cos\theta)
\end{equation}
where we have introduced
\begin{equation}
 \fl \label{15}
b(n,j,k,m):=
\frac{m!\,\Gamma(m+\frac{n-2}{2})\,\Gamma(j-k+\frac{n-2}{2})}
{k!\,(m-k)!\,\Gamma(\frac{n-2}{2})\,\Gamma(j+m-k+\frac{n}{2})}
\,\textstyle(j+m-2k+\frac{n-2}{2})\;.
\end{equation}
Replacing now the first Gegenbauer polynomial in \eref{12} with the help of
\eref{14} we realize that the integral \eref{12} represents in essence a
$3j $-symbol of the group $SO(2m+n)$, cf $(21)^1$:
\begin{equation}
\fl  \label{16}
\begin{array}{rl}
 G_n(j_1,j_2,j_3,m)=&\displaystyle
\frac{\Gamma(\frac{1}{2})\,\Gamma(m+\frac{n-1}{2})}{\Gamma(m+\frac{n}{2})\,
      [(2m+n-3)!]^3}
\sum_{k=0}^{\min\{m,[j_1/2]\}}(-1)^k\, b(n,j_1,k,m)\\[4mm]
&\hspace{-10mm}\displaystyle\times
\frac{(j_1-2k+2m+n-3)!\,(j_2+m+n-3)!\,(j_3+m+n-3)!}
     {(j_1-2k)!\,(j_2-m)!\,(j_3-m)!}\\[4mm]
&\displaystyle\times
\left(\begin{array}{ccc}
       j_1-2k &j_2-m &j_3-m\\ 0 & 0 & 0 
      \end{array}\right)_{(n+2m)}^2\;.
\end{array}
\end{equation}
Thus we have succeeded in expressing the integral \eref{12} by a single finite
sum and in turn found  a rather simple expression for the integral \eref{1}
in terms of four finite sums and $ 3j$-symbols of
$SO(2m+n)$ with $m=0,1,\ldots,\min\{\ell_4,\ell_5,\ell_6\}$. Expression
\eref{11} together with \eref{7}, \eref{15} and \eref{16} thus provides us
with an elementary formula for 
$ I_n(\ell_1,\ell_2,\ell_3 |\ell_4,\ell_5,\ell_6)$, which
is fairly simple and can easily be evaluated using, for example, some
computer-algebra program like {\it Mathematica}.\footnote{A {\it Mathematica}
 package which implements the results of this paper can be
  obtained from the authors upon request.} We also note that in our result
gamma functions with a half-integer argument always occur in terms of a
quotient and therefore $I_n$ is, for given integer $\ell$'s and $n$, a rational
number. 
\section{Discussion}
In this section we will briefly discuss the symmetry properties of
$I_n(\ell_1,\ell_2,\ell_3 |\ell_4,\ell_5,\ell_6)$, the corresponding
$6j $-symbol and its relation to the result of Ali\v{s}auskas
\cite{Sigitas}. First we note that the $3j$-symbol \eref{3} is obviously 
invariant under any permutation of the $ \ell$'s. In addition we note that
because of the first $ 3j$-symbol appearing on the right-hand side of \eref{2}
the phase factor in front of it may be replaced by
$(-1)^{\ell_1+\ell_2+\ell_3+\ell_4+\ell_5+\ell_6}$ as $ \ell_1+\ell_2+\ell_3$
is required to be an even integer. As a consequence 
$I_n(\ell_1,\ell_2,\ell_3 |\ell_4,\ell_5,\ell_6) $ and the $ 6j$-symbol have
identical symmetry properties. Using the invariance property of the Haar
measure in \eref{1} one easily verifies that 
\begin{equation}
  \fl\label{17}
  \begin{array}{r}
\left\{\begin{array}{ccc}
      \ell_1 &\ell_2 &\ell_3\\ \ell_4 &\ell_5 &\ell_6
      \end{array}\right\}_{(n)}
=
\left\{\begin{array}{ccc}
      \ell_2 &\ell_3 &\ell_1\\ \ell_5 &\ell_6 &\ell_4
      \end{array}\right\}_{(n)}
=
\left\{\begin{array}{ccc}
      \ell_3 &\ell_1 &\ell_2\\ \ell_6 &\ell_4 &\ell_5
      \end{array}\right\}_{(n)}
=
\left\{\begin{array}{ccc}
      \ell_2 &\ell_1 &\ell_3\\ \ell_5 &\ell_4 &\ell_6
      \end{array}\right\}_{(n)}
\\[4mm]
=
\left\{\begin{array}{ccc}
      \ell_1 &\ell_3 &\ell_2\\ \ell_4 &\ell_6 &\ell_5
      \end{array}\right\}_{(n)}
=
\left\{\begin{array}{ccc}
      \ell_3 &\ell_2 &\ell_1\\ \ell_6 &\ell_5 &\ell_4
      \end{array}\right\}_{(n)}
=
\left\{\begin{array}{ccc}
      \ell_1 &\ell_5 &\ell_6\\ \ell_4 &\ell_2 &\ell_3
      \end{array}\right\}_{(n)}\;.
  \end{array}
\end{equation}
These are indeed the well-known \cite{9} symmetries of the $ 6j$-symbols for
the group $ SO(3)$ which are thus shown to be valid for all $n\geq 3$ if
class-one representations are considered only. The
additional Regge symmetry \cite{10} known for the case $ n=3$ cannot be
verified by these methods and thus it is not clear whether it holds for
arbitrary $ n>3$. In combination with these symmetry properties Table 1
presents for all $\ell_i \in \{1,2,3,4\}$ explicit values for
$I_n(\ell_1,\ell_2,\ell_3|\ell_4,\ell_5,\ell_6)$ and  
\begin{equation}
  \label{18}
  c^{(\alpha)}_{\ell_1\ell_2\ell_3\ell_4\ell_5\ell_6}:=
  d_{\ell_1}\,d_{\ell_2}\,d_{\ell_3}\,d_{\ell_4}\, d_{\ell_5}\,d_{\ell_6}\,
  I_n(\ell_1,\ell_2,\ell_3|\ell_4,\ell_5,\ell_6)\;,
\end{equation}
where $d_\ell:=(2\ell+n-2)(\ell+n-3)!/\ell!(n-2)!$ denotes the dimension of the
$\ell$th representation.
Note that the quantity \eref{18} is the actual contribution of the
$\alpha$-topology to the high-temperature expansion of $SO(n)$-symmetric
lattice models \cite{4}. Thus with the tabulated quantities \eref{18} one can
derive high-temperature expansions for $SO(n)$-symmetric lattice models to
rather high order in the inverse temperature. For example, with only a few of
the tabulated values of \eref{18} one can find all expansion coefficients up
to order ten for the free energy \cite{11} of a mixed isovector-isotensor
model which recently has attracted much attention \cite{5,12}.

Finally we would like to comment on the relation of our result with that of
Ali\v{s}auskas \cite{Sigitas} on the $6j$-symbol. First we recall that with
our explicit result \eref{11} for $I_n$ we have with the help of \eref{2} a
similar representation for the $6j$-symbol, at least for those cases where the
additional $3j$-symbols appearing on the right-hand side of \eref{2} do not
vanish. Here the result has been derived via explicit group integration,
whereas Ali\v{s}auskas \cite{Sigitas} uses a series representation of the
$6j$-symbol in terms of isoscalar factors. Indeed, this representation
(equation (5.1) in \cite{Sigitas}) is very much similar in form to our result
\eref{11} for $I_n$. Note that the quantity $G_n$ defined in \eref{12} is in
fact closely related to an isoscalar factor of $SO(n)$, cf
$(41)^1$-$(44)^1$. In addition to that Ali\v{s}auskas \cite{Sigitas} was also
able to show that these isoscalars may be expressed in terms of (generalised)
$6j$-coefficients of $SU(2)$ which further allowed him to simplify his series
representation to three finite sums, see (5.7) in \cite{Sigitas} which is
valid for $n\geq 5$. 
In contrast to this, we have considered not the $6j$-symbol itself but
the group integral $I_n$ and represented it by four finite sums. As long as
the involved representation labels $\ell$ are small enough, which is actually
the case for a high-temperature expansion, this does not cause any
disadvantage. The advantage of considering
$I_n(\ell_1,\ell_2,\ell_3|\ell_4,\ell_5,\ell_6)$ respectively
$c^{(\alpha)}_{\ell_1\ell_2\ell_3\ell_4\ell_5\ell_6}$ is that the resulting
expressions (see Table 1) are valid for all $n\geq 2$ and thus allow for a
general discussion of the high-temperature expansion of $SO(n)$-symmetric
lattice models including the important $XY$-model ($n=2$) and Heisenberg model
($n=3$).  


\begin{table}
\caption{Explicit expressions for the non-vanishing integrals 
$I_n(\ell_1,\ell_2,\ell_3|\ell_4,\ell_5,\ell_6)$ defined in \eref{1} and the
corresponding quantity $c^{(\alpha)}_{\ell_1\ell_2\ell_3\ell_4\ell_5\ell_6}$
defined in \eref{18}.}
\begin{tabular}{@{}lll}
\br
$(\ell_1\ell_2\ell_3\ell_4\ell_5\ell_6)$ &
$I_n(\ell_1,\ell_2,\ell_3|\ell_4,\ell_5,\ell_6)$ &
$c^{(\alpha)}_{\ell_1\ell_2\ell_3\ell_4\ell_5\ell_6}$\\
\mr
$ ( {1\ 1\ 2\ 1\ 1\ 2} ) $ & $ \frac{4\ (n-2)}{(n-1)\ {n^3}\ {{(n+2)}^3}}$ & $
\frac{(n-2)\ (n-1)\ n}{n+2} $ \\[2mm] 
 
$ ( {1\ 1\ 2\ 1\ 3\ 2} ) $ & $ \frac{24}{{{(n-1)}^2}\ n\ {{(n+2)}^3}\ (n+4)}$
& $ \frac{(n-1)\ {n^3}}{n+2} $ \\[2mm]  
 
$ ( {1\ 1\ 2\ 2\ 2\ 1} ) $ & $ \frac{8\ (n-2)}{{{(n-1)}^2}\ {n^2}\
{{(n+2)}^3}}$ & $ (n-2)\ (n-1)\ n $ \\[2mm]  
 
$ ( {1\ 1\ 2\ 2\ 2\ 3} ) $ & $ \frac{48\ (n-2)}{{{(n-1)}^3}\ n\ {{(n+2)}^3}\
{{(n+4)}^2}}$ & $ \frac{(n-2)\ (n-1)\ {n^2}}{n+4} $ \\[2mm]  
 
$ ( {1\ 1\ 2\ 2\ 4\ 3} ) $ & $ \frac{288}{{{(n-1)}^3}\ n\ {{(n+2)}^2}\
{{(n+4)}^2}\ (n+6)}$ & $ \frac{(n-1)\ {n^3}\  (n+1 )}{2\ (n+4)} $ \\[2mm]  
 
$ ( {1\ 1\ 2\ 3\ 3\ 2} ) $ & $ \frac{72\ (n-2)\ (n+1)}{{{(n-1)}^3}\ {n^2}\
{{(n+2)}^3}\ {{(n+4)}^2}}$ & $ \frac{(n-2)\ (n-1)\ {n^2}\ (n+1)}{2\ (n+2)} $
\\[2mm]  
 
$ ( {1\ 1\ 2\ 3\ 3\ 4} ) $ & $ \frac{864\ (n-2)}{{{(n-1)}^3}\ {n^3}\ (n+2)\
{{(n+4)}^2}\ {{(n+6)}^2}}$ & $ \frac{(n-2)\ (n-1)\ {n^2}\ (n+1)}{2\ (n+6)} $
\\[2mm]  
 
$ ( {1\ 1\ 2\ 4\ 4\ 3} ) $ & $ \frac{1152\ (n-2)}{{{(n-1)}^3}\ {n^3}\ (n+1)\
{{(n+4)}^2}\ {{(n+6)}^2}}$ & $ \frac{(n-2)\ (n-1)\ {n^2}\ (n+1)\ (n+2)}
{6\ (n+4)} $ \\[2mm]  
 
$ ( {1\ 2\ 3\ 1\ 2\ 3} ) $ & $ \frac{72\ (n-2)}{{{(n-1)}^3}\ n\ {{(n+2)}^3}\
{{(n+4)}^3}}$ & $ \frac{(n-2)\ (n-1)\ {n^3}}{2\ (n+2)\ (n+4)} $ \\[2mm]  
 
$ ( {1\ 2\ 3\ 1\ 4\ 3} ) $ & $ \frac{864}{{{(n-1)}^3}\ {n^2}\ (n+2)\
  {{(n+4)}^3}\ (n+6)}$ & $ \frac{(n-1)\ {n^3}\  (n+1)}{2\ (n+4)} $ \\[2mm] 
 
$ ( {1\ 2\ 3\ 2\ 3\ 2} ) $ & $ \frac{288\ (n-2)\ (n+1)}{{{(n-1)}^4}\ n\
{{(n+2)}^3}\ {{(n+4)}^3}}$ & $ \frac{(n-2)\ (n-1)\ {n^2}\ (n+1)}{n+4} $
\\[2mm]  
 
$ ( {1\ 2\ 3\ 2\ 3\ 4} ) $ & $ \frac{1728\ (n-2)}{{{(n-1)}^4}\ n\ {{(n+2)}^2}\
{{(n+4)}^3}\ {{(n+6)}^2}}$ & $ \frac{(n-2)\ (n-1)\ {n^3}\ (n+1)}{2\ (n+4)\
(n+6)} $ \\[2mm]  
 
$ ( {1\ 2\ 3\ 3\ 2\ 3} ) $ & $ \frac{864\ (n-2)\ (n+1)}{{{(n-1)}^4}\ n\
{{(n+2)}^3}\ {{(n+4)}^3}\ (n+6)}$ & $ \frac{(n-2)\ (n-1)\ {n^3}\ (n+1)}{(n+2)\
(n+6)} $ \\[2mm]  
 
$ ( {1\ 2\ 3\ 3\ 4\ 3} ) $ & $ \frac{5184\ (n-2)}{{{(n-1)}^4}\ {n^2}\ (n+2)\
{{(n+4)}^3}\ {{(n+6)}^2}}$ & $ \frac{(n-2)\ (n-1)\ {n^3}\ (n+1)}{2\ (n+6)} $
\\[2mm]  
 
$ ( {1\ 2\ 3\ 4\ 3\ 2} ) $ & $ \frac{864\ (n-2)}{{{(n-1)}^4}\ n\ {{(n+2)}^2}\
{{(n+4)}^3}\ (n+6)}$ & $ \frac{(n-2)\ (n-1)\ {n^3}\ (n+1)}{4\ (n+4)} $ \\[2mm]

$ ( {1\ 2\ 3\ 4\ 3\ 4} ) $ & $ \frac{20736\ (n-2)}{{{(n-1)}^4}\ {n^2}\ (n+1)\
{{(n+4)}^3}\ {{(n+6)}^2}\ (n+8)}$ & $ \frac{(n-2)\ (n-1)\ {n^3}\ (n+1)\ (n+2)}
{2\ (n+4)\ (n+8)} $ \\[2mm]  
 
$ ( {1\ 3\ 4\ 1\ 3\ 4} ) $ & $ \frac{3456\ (n-2)}{{{(n-1)}^3}\ {n^3}\ (n+1)\
{{(n+4)}^3}\ {{(n+6)}^3}}$ & $ \frac{(n-2)\ (n-1)\ {n^3}\ (n+1)}{6\ (n+4)\
(n+6)} $ \\[2mm]  
 
$ ( {1\ 3\ 4\ 2\ 4\ 3} ) $ & $ \frac{20736\ (n-2)\ (n+2)}{{{(n-1)}^4}\ {n^3}\
(n+1)\ {{(n+4)}^3}\ {{(n+6)}^3}}$ & $ \frac{(n-2)\ (n-1)\ {n^2}\ (n+1)\
{{(n+2)}^2}}{2\ (n+4)\ (n+6)} $ \\[2mm]  
 
$ ( {1\ 3\ 4\ 3\ 3\ 4} ) $ & $ \frac{62208\ (n-2)}{{{(n-1)}^4}\ {n^2}\ (n+1)\ {{(n+4)}^3}\ {{(n+6)}^3}\ (n+8)}$ & $ \frac{(n-2)\ (n-1)\ {n^4}\ (n+1)}{2\ (n+6)\ (n+8)} $ \\[2mm] 
 
$ ( {1\ 3\ 4\ 4\ 4\ 3} ) $ & $ \frac{124416\ (n-2)\ (n+3)}{{{(n-1)}^4}\ {n^2}\
{{(n+1)}^2}\ {{(n+4)}^3}\ {{(n+6)}^3}\ (n+8)}$ & $ \frac{(n-2)\ (n-1)\ {n^4}\
(n+1)\ (n+3)}{4\ (n+4)\ (n+8)} $ \\[2mm]  
 
$ ( {2\ 2\ 2\ 2\ 2\ 2} ) $ & $ \frac{64\ (n-2)\  ({n^2}+4 n-24)}
{{{(n-1)}^5}\ {{(n+2)}^3}\ {{(n+4)}^3}}$ & $ \frac{(n-2)\ (n-1)\
{{(n+2)}^3}\  ({n^2}+4 n-24 )}{{{(n+4)}^3}} $ \\[2mm]  
 
$ ( {2\ 2\ 2\ 2\ 2\ 4} ) $ & $ \frac{768\ (n-2)\ n}{{{(n-1)}^5}\ {{(n+2)}^3}\
{{(n+4)}^3}\ (n+6)}$ & $ \frac{(n-2)\ (n-1)\ {n^2}\ (n+1)\
{{(n+2)}^2}}{{{(n+4)}^3}} $ \\[2mm]  
 
$ ( {2\ 2\ 2\ 2\ 4\ 4} ) $ & $ \frac{4608\ (n-2)}{{{(n-1)}^5}\ (n+1)\ (n+2)\
{{(n+4)}^3}\ {{(n+6)}^2}}$ & $ \frac{(n-2)\ (n-1)\ {n^2}\ (n+1)\
{{(n+2)}^3}}{2\ {{(n+4)}^3}} $ \\
\br
\end{tabular}
\end{table}
\setcounter{table}{0}
\begin{table}
\caption{\it cont.}
\begin{tabular}{@{}ll}
\br
$(\ell_1\ell_2\ell_3\ell_4\ell_5\ell_6)$ &
$I_n(\ell_1,\ell_2,\ell_3|\ell_4,\ell_5,\ell_6)$\\
\mr
$ ( {2\ 2\ 2\ 3\ 3\ 3} ) $ & $ \frac{864\ (n-2)\ (n+1)\ ({n^3}+8{n^2}-28n-48)}
{{{(n-1)}^5}\ {n^2}\ {{(n+2)}^3}\ {{(n+4)}^3}\ {{(n+6)}^2}}$ 
\\[2mm]

$ ( {2\ 2\ 2\ 4\ 4\ 4} ) $ & $ \frac{18432\ (n-2)\  ({n^3}+12 {n^2}-24 n-128)}
 {{{(n-1)}^5}\ {n^2}\ {{(n+1)}^2}\ {{(n+4)}^3}\ {{(n+6)}^2}\ {{(n+8)}^2}}$
 \\[2mm] 

$ ( {2\ 2\ 4\ 2\ 2\ 4} ) $ & $ \frac{2304\ (n-2)\ {n^2}}{{{(n-1)}^5}\ (n+1)\
{{(n+2)}^3}\ {{(n+4)}^3}\ {{(n+6)}^3}}$ \\[2mm] 

$ ( {2\ 2\ 4\ 2\ 4\ 4} ) $ & $ \frac{55296\ (n-2)}{{{(n-1)}^5}\ {{(n+1)}^2}\
{{(n+4)}^3}\ {{(n+6)}^3}\ (n+8)}$ \\[2mm] 

$ ( {2\ 2\ 4\ 3\ 3\ 3} ) $ & $ \frac{20736\ (n-2)}{{{(n-1)}^5}\ {{(n+2)}^2}\
{{(n+4)}^3}\ {{(n+6)}^3}} $ \\[2mm]  
 
$ ( {2\ 2\ 4\ 4\ 4\ 2} ) $ & $ \frac{13824\ (n-2)\ n\ (n+3)}{{{(n-1)}^5}\
{{(n+1)}^2}\ {{(n+2)}^2}\ {{(n+4)}^3}\ {{(n+6)}^3}} $ \\[2mm]  
 
$ ( {2\ 2\ 4\ 4\ 4\ 4} ) $ & $ \frac{663552\ (n-2)\ (n+3)}{{{(n-1)}^5}\
{{(n+1)}^3}\ {{(n+4)}^3}\ {{(n+6)}^3}\ {{(n+8)}^2}} $ \\[2mm]  
 
$ ( {2\ 3\ 3\ 2\ 3\ 3} ) $ & $ \frac{2592\ (n-2)\ (n+1)\  (2 {n^4}+17 {n^3}-14
{n^2}-84 n-72 )}
{{{(n-1)}^5}\ {n^3}\ {{(n+2)}^3}\ {{(n+4)}^3}\ {{(n+6)}^3}} $ \\[2mm] 
 
$ ( {2\ 3\ 3\ 3\ 4\ 4} ) $ & $ \frac{124416\ (n-2)\  ({n^3}+10 {n^2}-20n-48)}
{{{(n-1)}^5}\ {n^3}\ (n+1)\ {{(n+4)}^3}\ {{(n+6)}^3}\ {{(n+8)}^2}} $ \\[2mm] 
 
$ ( {2\ 3\ 3\ 4\ 3\ 3} ) $ & $ \frac{15552\ (n-2)\  ({n^3}+11 {n^2}-48 n-36 )}
{{{(n-1)}^5}\ {n^3}\ (n+2)\ {{(n+4)}^3}\ {{(n+6)}^3}\ (n+8)} $ \\[2mm] 
 
$ ( {2\ 4\ 4\ 2\ 4\ 4} ) $ & $ \frac{221184\ (n-2)\ (n+2)\  
(3 {n^4}+40 {n^3}+72 {n^2}-192 n-512 )}
{{{(n-1)}^5}\ {n^3}\ {{(n+1)}^3}\ {{(n+4)}^3}\ {{(n+6)}^3}\ {{(n+8)}^3}} $
\\[2mm]  
 
$ ( {2\ 4\ 4\ 4\ 4\ 4} ) $ & $ \frac{3981312\ (n-2)\ (n+2)\ (n+3)\  
({n^3}+14 {n^2}-16 n-128)}
{{{(n-1)}^5}\ {n^2}\ {{(n+1)}^4}\ {{(n+4)}^3}\ {{(n+6)}^3}\ {{(n+8)}^3}\
(n+10)} $ \\[2mm]  
 
$ ( {3\ 3\ 4\ 3\ 3\ 4} ) $ & $ \frac{186624\ (n-2)\  (4 {n^2}+37 n-50 )}
{{{(n-1)}^5}\ {n^2}\ (n+1)\ {{(n+4)}^3}\ {{(n+6)}^3}\ {{(n+8)}^3}} $ \\[2mm] 
 
$ ( {3\ 3\ 4\ 4\ 4\ 3} ) $ & $ \frac{373248\ (n-4)\ (n-2)\ (n+3)\ (n+20)}
{{{(n-1)}^5}\ {n^2}\ {{(n+1)}^2}\ {{(n+4)}^3}\ {{(n+6)}^3}\ {{(n+8)}^3}} $
\\[2mm]  
 
$ ( {4\ 4\ 4\ 4\ 4\ 4} ) $ & $ \frac{11943936\ (n-2)\ (n+3)\  
({n^6}+43{n^5}+400 {n^4}- 212 {n^3}-6752 {n^2}-5888 n+15360 )}
{{{(n-1)}^5}\ {n^2}\ {{(n+1)}^5}\ {{(n+4)}^3}\ {{(n+6)}^3}\ {{(n+8)}^3}\
  {{(n+10)}^3}} $  \\
\br
\end{tabular}
\end{table}
\setcounter{table}{0}
\begin{table}
\caption{\it cont.}
\begin{tabular}{@{}ll}
\br
$(\ell_1\ell_2\ell_3\ell_4\ell_5\ell_6)$ &
$c^{(\alpha)}_{\ell_1\ell_2\ell_3\ell_4\ell_5\ell_6}$\\
\mr
$ ( {2\ 2\ 2\ 3\ 3\ 3} ) $ 
& $\frac{(n-2)\ (n-1)\ n\ (n+1)\ ({n^3}+8 {n^2}-28 n- 48)}{2\ {{(n+6)}^2}}
$\\[2mm]

$ ( {2\ 2\ 2\ 4\ 4\ 4} ) $  & $ \frac{(n-2)\ (n-1)\ n\ (n+1)\ {{(n+2)}^3}\
(n+6)\  ({n^3}+12 {n^2}-24 n-128)}{ 6\ {{(n+4)}^3}\
{{(n+8)}^2}} $ \\[2mm]

$ ( {2\ 2\ 4\ 2\ 2\ 4} ) $ & $ \frac{(n-2)\ (n-1)\ {n^4}\ (n+1)\ (n+2)}
{4\ {{(n+4)}^3}\ (n+6)} $ \\[2mm] 

$ ( {2\ 2\ 4\ 2\ 4\ 4} ) $  & $ \frac{(n-2)\ (n-1)\ {n^3}\ (n+1)\
{{(n+2)}^3}}{2\ {{(n+4)}^3}\ (n+8)} $ \\[2mm] 

$ ( {2\ 2\ 4\ 3\ 3\ 3} ) $ & $ \frac{(n-2)\ (n-1)\ {n^4}\ (n+1)}{{{(n+6)}^2}}
$ \\[2mm]  
 
$ ( {2\ 2\ 4\ 4\ 4\ 2} ) $ & $ \frac{(n-2)\ (n-1)\ {n^4}\ (n+1)\ (n+2)\
(n+3)}{8\ {{(n+4)}^3}} $ \\[2mm]  
 
$ ( {2\ 2\ 4\ 4\ 4\ 4} ) $ & $ \frac{(n-2)\ (n-1)\ {n^4}\ (n+1)\ {{(n+2)}^2}\
(n+3)\ (n+6)}{2\ {{(n+4)}^3}\ {{(n+8)}^2}} $ \\[2mm]  
 
$ ( {2\ 3\ 3\ 2\ 3\ 3} ) $ & $ \frac{ (n-2)\ (n-1)\ n\ (n+1)\ (n+4)\
 (2 {n^4}+17 {n^3}-14 {n^2}-84 n-72)}{ 2\ (n+2)\ {{(n+6)}^3}}$ \\[2mm] 
 
$ ( {2\ 3\ 3\ 3\ 4\ 4} ) $ & $ \frac{(n-2)\ (n-1)\ {n^2}\ (n+1)\ (n+2)\  
({n^3}+10 {n^2}-20 n-48)}{2\ (n+6)\ {{(n+8)}^2}} $ \\[2mm] 
 
$ ( {2\ 3\ 3\ 4\ 3\ 3} ) $ & $ \frac{(n-2)\ (n-1)\ {n^2}\ (n+1)\ (n+4)\ 
({n^3}+11 {n^2}-48 n-36)}{4\ {{(n+6)}^2}\ (n+8)} $ \\[2mm] 
 
$ ( {2\ 4\ 4\ 2\ 4\ 4} ) $ & $ \frac{(n-2)\ (n-1)\ n\ (n+1)\ {{(n+2)}^3}\
(n+6)\ (3 {n^4}+40 {n^3}+72 {n^2}-192 n-512)}
{6\ {{(n+4)}^3}\ {{(n+8)}^3}} $\\[2mm] 
 
$ ( {2\ 4\ 4\ 4\ 4\ 4} ) $ & $ \frac{(n-2)\ (n-1)\ {n^3}\ (n+1)\ {{(n+2)}^2}\
(n+3)\ {{(n+6)}^2}\ ({n^3}+14 {n^2}-16 n-128)}
{4\ {{(n+4)}^3}\ {{(n+8)}^3}\ (n+10)}$ \\[2mm] 
 
$ ( {3\ 3\ 4\ 3\ 3\ 4} ) $ & $ \frac{(n-2)\ (n-1)\ {n^4}\ (n+1)\ (n+4)\ 
(4 {n^2}+37 n-50)}{4\ (n+6)\ {{(n+8)}^3}} $ \\[2mm] 
 
$ ( {3\ 3\ 4\ 4\ 4\ 3} ) $ & $ \frac{(n-4)\ (n-2)\ (n-1)\ {n^4}\ (n+1)\ (n+3)\ (n+20)}{8\ {{(n+8)}^3}} $ \\[2mm] 
 
$ ( {4\ 4\ 4\ 4\ 4\ 4} ) $ & $  \frac{(n-2)\ (n-1)\ {n^4}\ (n+1)\ (n+3)\
{{(n+6)}^3}\ 
({n^6}+43 {n^5}+ 400 {n^4}-212 {n^3}-6752 {n^2}-5888 n+15360)} 
{ 16\ {{(n+4)}^3}\ {{(n+8)}^3}\ {{(n+10)}^3}}$ \\[2mm] 
\br
\end{tabular}
\end{table}

\end{document}